\documentclass[aps,prl,twocolumn]{revtex4-2}

\usepackage[utf8]{inputenc}
\usepackage{amsmath,amsfonts}
\usepackage{graphicx}
\usepackage{textcomp}
\usepackage{xr}
\usepackage[nodisplayskipstretch]{setspace}
\newcommand{\figlet}[1]{{\sffamily #1}}

\newcommand{\Dmain}{1.8 \pm 0.1}
\newcommand{\Dsoft}{5.5 \pm 0.1}
\newcommand{\Dratio}{3.1}
\newcommand{\Doverrho}{0.97 \pm 0.06}

\begin{document}
%%%%%%%%%%%%%%%%%%%%%%%%%%%%%%%%%%%%%%%%%%%%

\title{A granular B{\"u}ttiker-Landauer motor}

\date{\today}

\author{O.~Devauchelle}
\email[]{devauchelle@ipgp.fr}
\affiliation{Université de Paris, Institut de Physique du Globe de Paris, 1 rue Jussieu, CNRS, 75238 Paris, France}

\author{P.~Popović}
\affiliation{Faculty of Physics, University of Belgrade, POB 44, 11001 Belgrade, Serbia}

\author{P.~Szymczak}
\affiliation{Institute of Theoretical Physics, Faculty of Physics, University of Warsaw, Pasteura 5, 02-093 Warsaw, Poland}

\author{A.~Abramian}
\affiliation{Sorbonne Université, CNRS, Institut Jean Le Rond d'Alembert, F-75005 Paris, France}

\author{A.~Lazarus}
\affiliation{Sorbonne Université, CNRS, Institut Jean Le Rond d'Alembert, F-75005 Paris, France}
\affiliation{Massachusetts Institute of Technology, Department of Mathematics, Cambridge, MA 02139, USA}

\begin{abstract}
Random walkers usually diffuse according to Fick's law. On average, they move down the gradient of their concentration and, in the absence of external force, tend to distribute themselves uniformly. In some experiments, however, this familiar notion is at odds with observation. Sand grains, for instance, gather along the nodal lines of a vibrated elastic plate to form a Chladni figure, thus accumulating where fluctuations are weak---a fact that escapes the reach of Fick's law. On theoretical grounds,
Büttiker [\emph{Zeitschrift für Physik B}, {\bf 68}, 1987] %\citet{buttiker1987transport}
and
Landauer [\emph{J Stat Phys}, {\bf 53}, 1988] %\citet{landauer1988motion}
proposed that particles submitted to a non-uniform temperature field would indeed gather where the temperature is low. They also predicted that, in the presence of a potential force, a non-uniform temperature could drive a steady current of particles, powered only by noise. Here, we present an experimental realization of these phenomena in a macroscopic system, which confirms the quantitative predictions of Büttiker.
\end{abstract}

%%%%%%%%%%%%%%%%%%%%%%%%%%%%%%%%%%%%%%%%%%%%
\maketitle
%%%%%%%%%%%%%%%%%%%%%%%%%%%%%%%%%%%%%%%%%%%%

In 1791, Ernst Chladni started touring Europe to show his mesmerizing experiment \cite{chladni1787entdeckungen, ullmann2007life}. He would vibrate a metal plate with a violin bow, and sprinkle sand over it. Within a few seconds, the sand would gather along the nodal lines of the vibration, thus visualizing acoustic modes.

Chladni was after a consistent theory of acoustics; the shape of the figure was the mystery, and the grains' role was only to draw it. It probably seemed obvious that the grains would settle along the quiet nodal lines. This notion is so intuitive, in fact, that one has to wait until the recent tide of active-matter research to see it investigated in details \cite{van2010inversion,arango2016stochastic}. According to these recent contributions, the pattern results from an effective average force which pushes the grains toward the nodes. This force can result from the accumulation of momentum during a vibration cycle \cite{kudrolli2008swarming}, much like coffee grains slide over the surface of a vibrating espresso machine, or from air currents induced by the vibration, as first suggested by \citet{faraday1831xvii}. In these models, fluctuations are unnecessary.

Based on experiments, \citet{grabec2017vibration} first suggested a purely statistical explanation, wherein each grain behaves like a random walker, whose steps shorten as it nears a nodal line. \citet{grabec2023diffusion} numerically confirmed that a collection of such walkers can form a Chladni figure. Surprisingly, however, Fick's law cannot account for this observation, at least in its canonical form ($\mathbf{j} = - D \nabla \rho \, ,$ where $\mathbf{j}$ is the flux of grains, $D$ their diffusivity, and $\rho$ their surface density). At equilibrium, indeed, when the flux $\mathbf{j}$ vanishes, the density $\rho$ must be uniform---even when $D$ is not.

To reconcile theory with intuition, \citet{abramian2024} proposed to use the diffusion law of a Knudsen gas in a heated pipe \cite{van1988explicit}: \begin{equation}
  \mathbf{j} = - \nabla \left( D \rho \right)
  \label{eq:newFick}
\end{equation}
instead of Fick's law. This expression has been a matter of theoretical debate for some time \cite{landauer1988motion, van1988relative}---a debate later revived in the context of chemotaxis \cite{schnitzer1993theory,tailleur2008statistical}, and related to the problem of Itô vs. Stratonovich integrals. In short, Eq.~\eqref{eq:newFick} does not hold for every system, but one can argue in its favor for bouncing grains \cite{hanggi1982nonlinear,van1988explicit,volpe2016effective}: One needs to assume that the grains are independent random walkers, whose step length and duration are set by the grain's location \emph{at the onset} of a jump. We present a detailed analysis of the microscopic behavior of bouncing grains in a companion paper \cite{companion}.

Intuitively, one can make sense of Eq.~\eqref{eq:newFick} by numerically simulating a random walk with space-dependent diffusivity (this requires only 8 lines of code, SM1). After the system has settled, we expect that the grain density does not change. Eq.~\eqref{eq:newFick} then requires that the grains distribute themselves over the vibrating plate in inverse proportion to diffusivity:
\begin{equation}
  \rho \propto 1/D \, ,
  \label{eq:one_over_D}
\end{equation}
which accords with numerical simulations (SM1) \cite{volpe2016effective} and Chladni experiments alike \cite{abramian2024}.

While apparently innocuous, the substitution of Eq.~\eqref{eq:newFick} for Fick's law has profound consequences on the macroscopic behavior of the walkers. For instance, \citet{landauer1988motion} noted that Eq.~\eqref{eq:one_over_D}, a steady-state solution, differs from the Boltzmann distribution. A system that obeys it thus cannot be in thermodynamic equilibrium.

\citet{buttiker1987transport} furthered this idea by (theoretically) submitting such a system to a potential force, and found that a constant current of particles should appear, with the particles drifting endlessly along steady streamlines. Surprisingly, this Brownian motor does not need a pawl \cite{reimann2002brownian,benjamin2008inertial}, unlike its forebear the Feynman-Smoluchowski ratchet \cite{smoluchowski1912experimentell}. To our knowledge, such a motor has never been realized experimentally (although it is sometimes seen as a simplified model of the Peltier-Seebeck effect \cite{reimann2002brownian}). Here, we build a macroscopic setup inspired by Chladni's experiment to do so, and to investigate the most striking consequences of the peculiar statistics of bouncing grains.

\begin{figure}
\includegraphics[width=\linewidth]{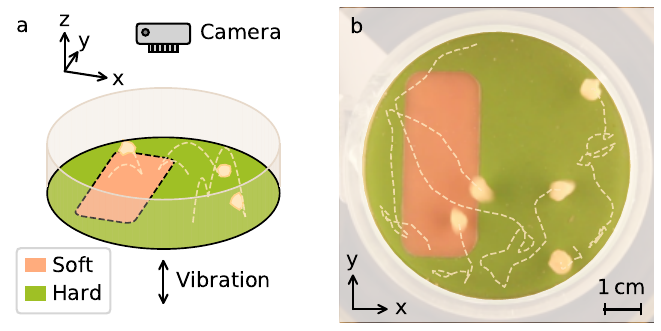}%
\caption{
\figlet{a}: Experimental setup. The disk vibrates along the $z$-axis. The setup can be tilted by a few degrees around the $x$-axis. Quartz grains bounce randomly over a hard plastic surface (green, PVC), or a soft silicon elastomer (pink, vinyl polysiloxane).
\figlet{b}: Still frame from an experimental movie (run C). Dashed white lines show the grains' trajectories during $1.5\,$s (75 frames).\label{fig:setup}}%
\end{figure}

For simplicity, we replace Chladni's elastic plate with a rigid disk, over which  grains can bounce and diffuse (radius $R=3.9\,$cm, PVC, green disk on Fig.~\ref{fig:setup}, experimental details in SM2). To create a diffusivity contrast, we cut a rectangular groove into the disk (area $10.3\,$cm$^2$), and fill it with a soft elastomer ($1\,$mm-thick vinyl polysiloxane, pink area on Fig.~\ref{fig:setup}). By shining a laser sheet at grazing incidence on the disk as we pour the elastomer, we ensure that its surface is level with that of the disk ($\pm10\,$\textmu m). Finally, we cover the entire surface of the disk with Parafilm M to reduce the build-up of static electricity, and enclose the disk with a $2\,$cm-high, transparent ring (PMMA). This makes up the corral where the bouncing grains will roam.

The corral is fixed atop a vertical aluminum rod attached with ball bearings to a steel rail, also vertical. The lower end of the rod is connected to the membrane of a bass loudspeaker, driven by a function generator through a $40\,$W amplifier. This setup allows us to shake the corral along the direction perpendicular to its surface ($z$ axis, Fig.~\ref{fig:setup}\figlet{a}) at a frequency $\nu$ between 30 and $40\,$Hz, and an amplitude $A$ up to a few millimeters. In practice, the heterogeneity of the substrate most matters near the threshold of bouncing \cite{abramian2024}, when $A(2\pi \nu)^2\sim g$ where $g$ is the acceleration of gravity, hence our choice of frequency.

A camera, held $33\,$cm above the disk, records the horizontal motion of the grains at a rate of $50\,$FPS. The entire setup, including the camera, rotates around the horizontal $x$ axis, so that we can use gravity to add a force in the plane of the vibrating disk. The inclination of the plate is measured with a capacitive inclinometer (Meiri ME 26410, precision $0.01^{\circ}$), and controlled by a micrometer screw driven by a step motor.

We begin our series of experiments with a horizontal corral, into which we drop $N=5$ grains of coarse quartz sand (diameter $d_s = 4.5\pm 0.5\,$mm, experimental run C, table~SM2). We then switch on the vibration and manually adjust its amplitude so the grains bounce over the disk up to about $0.5\,$cm above its surface. As they jump and rebound, the grains roam erratically across the corral (Fig.~\ref{fig:setup}\figlet{b}). We locate the grains based on the saturation of their color, which contrasts with that of the background, and track them to reconstruct their trajectories (dashed white lines in Fig.~\ref{fig:setup}\figlet{b}) \cite{munkres1957algorithms}.

\begin{figure}
\includegraphics[width=\linewidth]{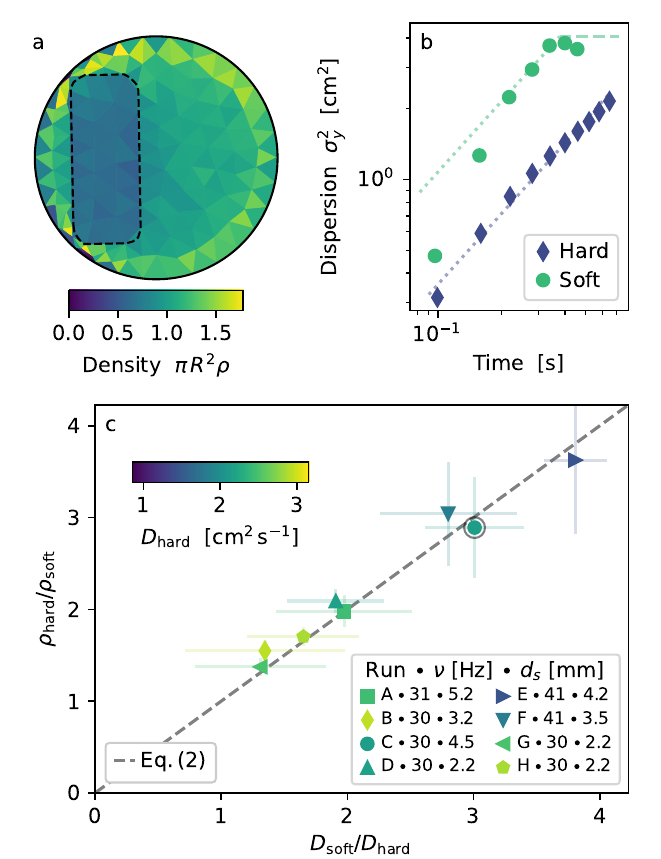}%
\caption{Steady state in a horizontal corral.
\figlet{a}: Surface density of grains for run C. The tilt of the setup is less than $0.5^{\circ}$.
\figlet{b}: Variance of the grains' trajectories along $y$, as a function of time (symbols). Dotted lines show linear fits. Dashed line indicates the width of the soft pad.
\figlet{c}: Ratio of density over the hard disk to density over the soft pad, as a function of the inverted ratio of diffusivities (symbols). Points with same $\nu$ and $d_s$ differ by plate acceleration (not measured). Color shows the diffusivity over the hard disk. Vertical error bars show the standard deviations over at least 6 movies. Horizontal error bars indicate the difference between dispersion and autocorrelation. Dashed gray line shows identity.
\label{fig:one_over_D}}%
\end{figure}

At first sight, the grains seem indifferent to the substrate, but a careful observer notes that they seem to avoid the soft pad. A more careful analysis of the trajectories confirms this impression. Indeed, collecting the trajectories of 15 one-minute-long movies, we can estimate their density distribution over the coral (Fig.~\ref{fig:one_over_D}\figlet{a}). We find that the grain density is almost 3 times larger over the hard surface than over the soft pad, while their distribution is relatively uniform within each domain (the density is a bit higher near the wall, perhaps due to shorter jumps). Eq.~\eqref{eq:one_over_D} suggests that this density difference might be associated with a diffusivity contrast.

To test this interpretation, we cut the trajectories where they cross the boundary of the soft pad, and distribute the resulting chunks into two categories, ``soft'' and ``hard'', after the substrate over which the grain bounces. For each category, we then plot the average square displacement of the trajectory along the $y$-axis, $\sigma^2_y$, as a function of time (Fig.~\ref{fig:one_over_D}\figlet{b}). For hard trajectories, we find that after about $0.1\,$s the average square displacement grows linearly with time---the signature of diffusion. The slope of this relation is twice the diffusivity, which we find to be $D_{\mathrm{hard}}\approx \Dmain \,$cm$^2\,$s$^{-1}$ for run C  (table~SM2). Over the soft pad, unfortunately, the signature of diffusion is less legible, because the transition between the ballistic and diffusive regimes occurs barely before the grain leaves the soft domain. In other words, the correlation length of the trajectory is barely shorter than the size of the soft pad (despite our choice of the $y$ direction). If, nonetheless, we adjust the diffusivity so that the linear relation is tangent to the actual dispersion curve, we find $D_{\mathrm{soft}}\approx \Dsoft \,$cm$^2\,$s$^{-1}$ for run C (this value is consistent with independent experiments \cite{companion}). We interpret this difference as follows: The grains bounce over the soft elastomer without dissipating much energy, which translates into faster and longer jumps. The ratio of the two diffusivities is  $D_{\mathrm{soft}}/D_{\mathrm{hard}}\approx \Dratio$, close enough to the inverse ratio of densities.

We repeat the experiment with grains of different sizes, and vibrate the coral at different frequencies (we also change the vibration amplitude, but we do not measure it). We then measure the grains' density and diffusivity, and plot the ratio of the former as a function of the inverse ratio of the latter (Fig.~\ref{fig:one_over_D}\figlet{c}). Despite large measurement uncertainties, the data points gather along the identity line, in accordance with Eq.~\eqref{eq:one_over_D}. The ratio of these two ratios is, on average over our 8 experiments, $ \Doverrho $ (standard deviation), thus supporting the use of Eq.~\eqref{eq:newFick}.

As long as the corral remains horizontal, no average force acts on the grains, and the heterogeneous distribution of Fig.~\ref{fig:one_over_D}\figlet{c} is therefore of purely statistical origin. That the same distribution appears on the vibrated membrane of \citet{abramian2024} suggests that the Chladni figure results from the same phenomenon.

In the present setup, however, we can apply a controlled force on the grains, by tilting the corral about the $x$ axis. According to B{\"u}ttiker's reasoning, doing so should induce a steady average current of particles, driven by fluctuations. In the present notations, indeed, the potential force enters Eq.~\eqref{eq:newFick} as
\begin{equation}
  \mathbf{j}/(\rho D) = -  \nabla \ln ( \rho D ) - ({\mu}/{D}) \nabla V \, ,
  \label{eq:log_rho_D}
\end{equation}
where $V$ is the gravity potential, and $\mu$ is a grain's mobility (the propensity of a particle to move in response to a force). Hereafter, we call ``temperature'' the ratio of diffusivity to mobility, and define its inverse as $\beta = \mu/D$.  When the temperature gradient is not aligned with the external force, the curl of the right-hand side cannot vanish, and there must be a finite current $\mathbf{j}$. Unfortunately, this average current is drowned in fluctuations, and we need to work out its theoretical value before we can identify it in the experiment. 

\begin{figure}
\includegraphics[width=\linewidth]{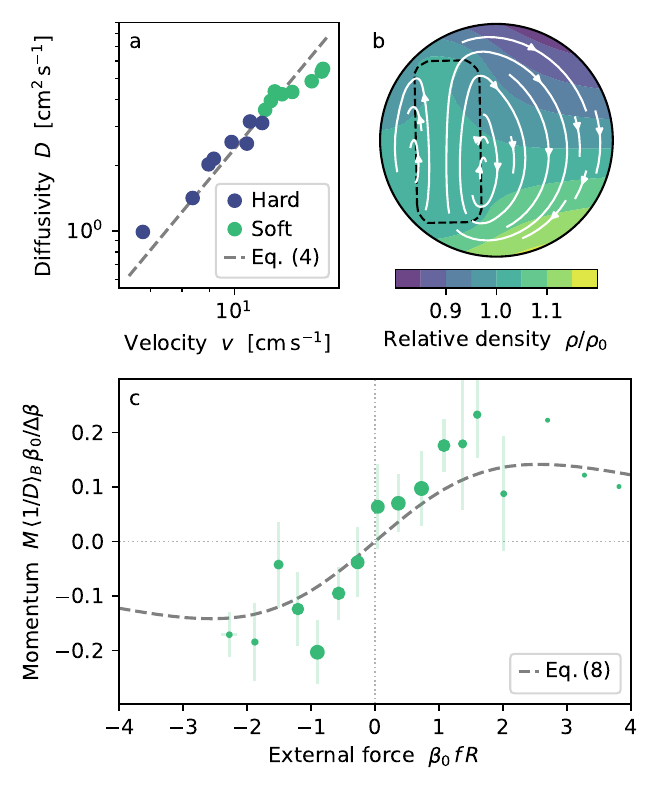}%
\caption{Non-equilibrium steady state with a particle current in a tilted corral.
\figlet{a}: Relation between diffusivity and in-plane root mean square velocity. Dots show average over all runs.
\figlet{b}: Finite-elements simulation of the steady-state Smoluchowski equation \eqref{eq:fe} for a small temperature variation ($\beta_0fR = 0.5$ and $\Delta \beta/\beta_0=1$). Contours: ratio of the grain density $\rho$ to the density of the base state, $\rho_0$. White streamlines show the grain current.
\figlet{c}: Normalized momentum as a function of normalized tilt for all experimental runs, distributed into 20 bins. The area of each dot is proportional to the number of movies in that bin (from 2 to 39). Error bars show standard deviation. Grey dashed line: finite-elements solution of the Smoluchowski equation, Eq.~\eqref{eq:M_fe}.
\label{fig:momentum}}%
\end{figure}

To estimate the inverse temperature $\beta$ of the bouncing grains, we represent the vertical trajectories of the bouncing grains as a series of ballistic parabolas (Fig.~\ref{fig:setup}\figlet{a}), which are set by the angle they form with the surface of the disk, $\theta$, and their velocity parallel to the surface of the disk, $v$. We further assume that the correlation time of the random walk is the duration of a jump, $\tau = (2v\tan \theta)/g$, that the horizontal bouncing direction is random, and that the grains do not interact (the probability of collision is less than 10$\,\%$ per jump, SM2D). Their diffusivity in two dimensions is then
\begin{equation}
  D = \tau v^2/4 = v^3 \tan \theta /(2g).
  \label{eq:D_v_theta}
\end{equation}
This relation matches our observations (Fig.~\ref{fig:momentum}\figlet{a}) if we identify $v$ with the root mean square velocity in the corral's plane, which we can measure straightforwardly from experimental trajectories. Fitting this relation to our observations, we find a bouncing angle of $\theta = 78^{\circ} \pm 4^{\circ}$. This reasoning also yields the mobility which, for a ballistic jump, reads  $\mu=\tau/(2m)$. The temperature is then
\begin{equation}
  {1}/{\beta}
    ={m v^2}/{2} \, ,
    \label{eq:fluct_diss}
\end{equation}
which is just the kinetic energy of a grain in the plane of the disk---a consistency check. In the companion paper, we develop the analogy with classical thermodynamics~\cite{companion}.

In steady state ($\partial \rho/\partial t = 0$), the balance of grains dictates that their average current be divergence free:
\begin{equation}
  \nabla \cdot \mathbf{j} = 0 \, .
  \label{eq:FP}
\end{equation}
In combination with Eq.~\eqref{eq:log_rho_D}, this yields the Smoluchowski (or Fokker-Planck) equation we need to solve. It depends on the non-uniform parameter $\beta$, and we could solve it numerically for the temperature distribution we measure in the experiments. This, however, would be cumbersome, and we prefer to derive a simplified version of it, in the hope that it will be more telling.

We assume that the temperature is almost uniform, that is $\Delta \beta \ll  \beta_0 $, where  $\Delta \beta \equiv \beta_{\mathrm{hard}} - \beta_{\mathrm{soft}}$ is the inverse-temperature difference between the hard disk and the soft pad, and $\beta_0=( \beta_{\mathrm{hard}} + \beta_{\mathrm{soft}} )/2$ is the average temperature. Then, the grains distribute themselves in the corral according to the Boltzmann distribution $\rho_0$, which we generalize for heterogeneous diffusivity $D$ while keeping the temperature $\beta_0$ constant:
$
  \rho_0 D =  {e^{-\beta_0 V}}/Z_0
$, where $Z_0$ is the partition function, which ensures that the integral of $\rho_0$ over the corral is one. This base state, which holds whenever the temperature is uniform, features no current of particles.

At the next order, we define the perturbation of the grain density $\phi$ such that
${\rho}/{\rho_0} =  1 +  \phi \,{\Delta \beta}/{\beta_0}$,
and we find that, in steady state, the Smoluchowski equations~\eqref{eq:log_rho_D} and \eqref{eq:FP} yield
\begin{equation}
  \nabla \cdot \left[  e^{-\beta_0 V} \left( \nabla \phi + H \beta_0 \nabla V \right) \right] = 0
  \label{eq:fe}
\end{equation}
where $H$ encodes the distribution of temperature ($H = 1/2$ over the hard disk, and $-1/2$ otherwise). We also require that no current crosses the wall. When the corral is tilted, gravity applies a constant force $f$ on the grains, which is directed along the $y$ axis. The resulting potential is $V = f y$ and, using the disk's radius $R$ as the unit of length, Eq.~\eqref{eq:fe} then depends on a single parameter, $\beta_0 f R$. The finite element method is well suited to solve this linear, elliptic equation (SM3) \citep{MR3043640}.

Fig.~\ref{fig:momentum}\figlet{b} shows an example of numerical solution. With respect to the base distribution $\rho_0$, the grains tend to concentrate in the lower part of the corral (negative $y$), but this concentration is stronger over the hard disk than over the soft pad, as if the grains did their best to match the Boltzmann distribution over each domain. Of course, they can only fail to do so, and their  attempt creates an excess concentration over the higher part of the (hotter) soft pad, which diffusion turns into a flux towards the comparatively depleted hard disk. Just as \citet{buttiker1987transport} foresaw, this generates a steady current of particles (white streamlines in Fig.~\ref{fig:momentum}\figlet{b}).

This current induces an angular momentum,  $M \equiv \iint  \mathbf{x} \times \mathbf{j} \, $, which derives straightforwardly from $\phi$ upon integration over the corral:
\begin{equation}
  M \dfrac{\beta_0}{\Delta \beta } \left< \dfrac{1}{D} \right>_B =
    - \left<  
      H \beta_0 f x
      + \mathbf{x} \times \nabla \phi
      \right>_B
    \, ,
  \label{eq:M_fe}
\end{equation}
where $\langle \cdot \rangle_B$ denotes a space average weighted by the Boltzmann factor, $\exp(-\beta_0 f y)$. Conveniently, this expression provides a scale for the momentum $M$, and thus allows us to plot it as a function of the only remaining dimensionless parameter, $\beta_0 f R$, for all experiments (Fig.~\ref{fig:momentum}\figlet{c} and SM4). Despite considerable scatter, the bin-averaged data points gather around the curve that represents Eq.~\eqref{eq:M_fe}, without any fitting. In particular, the angular momentum appears to saturate when the gravity potential gets strong enough to oppose temperature ($\beta_0 f R \sim 1$)---as one would expect.

In conclusion, the experiments reported here confirm that bouncing grains tend to accumulate where their diffusivity is weak, in the absence of any average force, and in contrast with Fick's law. A simple random walk can thus translate the vibration of an elastic plate into the Chladni figure, as first suggested by \citet{grabec2017vibration} and recently verified experimentally \cite{abramian2024}.

The combination of a temperature gradient with an external force  can generate a non-equilibrium steady state with a stationary current of particles---a rudimentary heat engine \cite{benjamin2008inertial}. Some feature of this engine are reminiscent of natural convection: Hot particles rise up, and then settle down through cold areas, thus transporting heat (Fig.~\ref{fig:momentum}\figlet{b}). Unlike natural convection, however, the present phenomenon requires no interaction between particles, and requires that the temperature be not aligned to gravity.

The gathering of the bouncing grains in places of lesser diffusivity is also reminiscent, albeit more distantly, of the tendency of some out-of-equilibrium systems to spend more time in highly dissipative configurations \cite{PhysRevLett.119.038001}. It seems reasonable to assume that, in the present experiment, the grains jump higher where the restitution coefficient is larger (on the bouncy silicon pad). As a consequence of heterogeneous diffusivity, the grains thus gather where restitution is poor, that is, where the dissipation of the grains' energy is stronger. The companion paper proposes a tentative theoretical treatment the thermodynamics of such systems \cite{companion}.

Altogether, the recipes reported here are rather easy. A random walker whose pace is reset at the beginning of each step suffices to break Fick's law. As for the Büttiker-Landauer motor, it only requires a temperature gradient orthogonal to some external force. That these phenomena do not seem to occur all around us is a mystery to the authors. Do we need to look more carefully, or is there something truly specific to bouncing sand grains?

\bigskip

\begin{acknowledgments}
The idea of the paper arose during seminal discussions with S.~Protière. We are also grateful to E.~Lajeunesse and F.~Métivier for their help with the experiment, and to D.H.~Rothman, G.~Pucci, A.~Eddi, W.~Herreman, A.P.~Petroff, A.~Estevez-Torres, S.~Djambov, A.J.C.~Ladd and M.G.~Worster for inspiring discussions.

O.D was partially funded by PhysErosion ANR-22-CE30-0017.
\end{acknowledgments}

%%%%%%%%%%%%%%%%%%%%%%%%%%%%%%%%%%%%%%%%%%%%
\bibliography{./biblio_BL.bib}
%%%%%%%%%%%%%%%%%%%%%%%%%%%%%%%%%%%%%%%%%%%%

%%%%%%%%%%%%%%%%%%%%%%%%%%%%%%%%%%%%%%%%%%%%
\end{document}